\documentclass[aps,pre,showpacs]{revtex4}

\usepackage{graphicx}

\renewcommand{\epsilon}{\varepsilon}

\newcommand{\figurewidth}{0.48\textwidth}
\newcommand{\widefigurewidth}{0.60\textwidth}

\begin{document}

\title{Colloidal stabilization via nanoparticle haloing}
\author{Jiwen Liu}
\author{Erik Luijten}
\email[Corresponding author. E-mail: ]{luijten@uiuc.edu}
\affiliation{Department of Materials Science and Engineering and
Frederick Seitz Materials Research Laboratory, University of Illinois at
Urbana-Champaign, Urbana, Illinois 61801}

\date{\today}

\begin{abstract}
  We present a detailed numerical study of effective interactions
  between micron-sized silica spheres, induced by highly charged
  zirconia nanoparticles.  It is demonstrated that the effective
  interactions are consistent with a recently discovered mechanism for
  colloidal stabilization.  In accordance with the experimental
  observations, small nanoparticle concentrations induce an effective
  repulsion that counteracts the intrinsic van der Waals attraction
  between the colloids and thus stabilizes the suspension. At higher
  nanoparticle concentrations an attractive potential is recovered,
  resulting in reentrant gelation. Monte Carlo simulations of this
  highly size-asymmetric mixture are made possible by means of a
  geometric cluster Monte Carlo algorithm. A comparison is made to
  results obtained from the Ornstein--Zernike equations with the
  hypernetted-chain closure.
\end{abstract}

\pacs{61.20.Ja, 64.75.+g, 82.70.Dd}

\maketitle

\section{Introduction}
\label{sec:intro}

Colloidal suspensions are used in a wide range of applications, such as
coatings and drug carriers, and as a precursor for various advanced
materials, including colloidal crystals. In all these systems, the
\emph{stability} of the suspension plays a central role.  The van der
Waals attractions between the colloids---if not counteracted by some
appropriate mechanism---lead to their aggregation, typically resulting
in the formation of a gel. Conventional methods to mitigate the effect
of the van der Waals forces rely on charge stabilization or steric
stabilization~\cite{hunter01}. In charge stabilization, an electrostatic
repulsion between the colloids is induced, e.g., by an appropriate
choice of the pH\@. Steric stabilization frequently involves grafting
short polymer chains onto the colloidal surface, which prevents a close
approach of the colloidal surfaces.

Quite recently, an alternative strategy for colloidal stabilization has
been discovered~\cite{tohver01}. These experiments involve an aqueous
suspension of silica spheres with a diameter~$\sigma_{\text{micro}}$ of
approximately $0.6\mu\mathrm{m}$, to which charged zirconia
nanoparticles of much smaller diameter ($\sigma_{\text{nano}} =
6\mathrm{nm}$) are added. In the absence of nanoparticles, as well as in
the presence of very small concentrations of nanoparticles, the
microspheres exhibit a tendency to aggregate owing to their van der
Waals attraction.  Increasing the nanoparticle concentration
\emph{prevents} this aggregation and thus stabilizes the suspension. At
even higher zirconia concentrations, the aggregation behavior reappears,
leading to a ``window of stability'' in the nanoparticle concentration.
In order to ensure that the observed stability does not result from
direct electrostatic repulsion between the microspheres, the suspension
is kept at a very low pH, near the isoelectric point of silica.
Furthermore, it has been verified in Ref.~\cite{tohver01} that there is
no strong adsorption of nanoparticles on the silica surface, which would
lead to an effective surface charge accumulation on the colloids and
thus again result in electrostatic stabilization. Specifically, the
amount of zirconia associated with the microspheres was determined from
supernatant measurements employing inductively coupled plasma analysis
(indicating weak adsorption) as well as from scanning angle
reflectometry of a silica surface immersed in the nanoparticle solution
(indicating no detectable adsorption at all). We view the former
analysis as more representative, since it involves the actual silica
microspheres for which the reported phase behavior was determined.

Since there are only very few approaches for the stabilization of
colloidal suspensions, it is of clear importance to explore the
underlying mechanism of these observations. Not only may a new
stabilization technique be of practical significance, the binary mixture
investigated in Ref.~\cite{tohver01} also constitutes an interesting
model system for the exploration of depletion effects~\cite{asakura54}
in nonadditive systems~\cite{mendez00,roth01,louis02}.  Lewis and
co-workers~\cite{tohver01} ascribe the initial colloidal stabilization
to the formation of a ``halo'' of zirconia particles around the silica
colloids, arising from the strong electrostatic repulsion between
nanoparticles.  However, while consistent with zeta-potential
measurements that confirm a weak accumulation of nanoparticles near the
colloidal surface~\cite{tohver01}, such halo formation is at variance
with the observation of stabilization at zirconia volume fractions below
$10^{-3}$, where the average nanoparticle separation greatly exceeds the
electrostatic screening length~$\lambda$. Indeed, a rather high
concentration of nitric acid is required to reach the isoelectric point
of silica, resulting in $\lambda \approx 2\mathrm{nm}$~\cite{tohver01b},
only one third of the nanoparticle diameter. On the other hand,
explanation of the reentrant gelation representing the upper boundary of
the stability window appears more straightforward and is attributed in
Ref.~\cite{tohver01} to the regular depletion
attraction~\cite{asakura54} induced by the nanoparticles. While this can
not be excluded \emph{a priori}, it must be noted that nonadditivity
effects can significantly alter the nature of the depletion interaction.

In Ref.~\cite{liu04b} we have studied this system by means of canonical
Monte Carlo simulations. While particle-based simulations can explicitly
account for the fluctuation and correlation effects that are ignored in
many analytical methods, it is important to note that the extreme size
asymmetry $\alpha = \sigma_{\text{micro}}/\sigma_{\text{nano}} = 100$
places the current system out of reach for conventional Monte Carlo and
molecular dynamics simulations. This was resolved by means of a novel,
highly efficient Monte Carlo scheme~\cite{dress95,geomc,liu05a} that is
capable of simultaneously equilibrating interacting species of vastly
different sizes. In this article, we present a detailed technical
account of the calculations reported in Ref.~\cite{liu04b} and make a
comparison with results obtained in an integral-equation study of the
same system~\cite{chavez04,karanikas04,karanikas05}.

Understanding and prediction of phase behavior and stability of a
suspension relies on a fundamental knowledge of the \emph{effective}
forces between colloids~\cite{russell89}, which arise from a combination
of direct interactions and indirect interactions mediated through the
solvent and through other solute particles. It is this potential of mean
force that we determine in the simulations, as a function of
nanoparticle concentration. Our findings confirm the experimental
observations on a quantitative level and, in combination with the
corresponding nanoparticle distributions, clarify the physical mechanism
of the stabilization. Consequently, our computational approach also has
predictive capabilities, allowing the stability window to be tuned as a
function of nanoparticle charge and microsphere--nanoparticle size
asymmetry.

\section{Colloid--nanoparticle interactions}
\label{sec:interaction}

It is the purpose of our calculation to determine the effective
interaction~$V_{\text{micro}}^{\rm eff}(r)$ between a pair of colloidal
microspheres induced by the nanoparticles.  Calculation of
$V_{\text{micro}}^{\rm eff}(r)$ amounts to integrating out the degrees of
freedom of the nanoparticles in the partition
function~\cite{hill56,dijkstra99},
\begin{equation}
e^{-\beta [V_{\text{micro}}^{\rm eff}(r) + V_{\text{micro}}(r)]} =
\sum_{\{s_{\text{nano}}\}} e^{-\beta \mathcal{H}(r; \{\mathbf{r}\})} \;,
\label{eq:V_eff}
\end{equation}
where $\mathcal{H}(r; \{\mathbf{r}\})$ is the Hamiltonian describing a
system containing two microspheres at a separation~$r$ and a set of
nanoparticles with coordinates~$\{\mathbf{r}\}$. $V_{\text{micro}}(r)$
represents the direct microsphere pair potential and $\beta$ indicates
the inverse temperature~$(k_{\rm B}T)^{-1}$, with $k_{\rm B}$
Boltzmann's constant and $T$ the absolute temperature. The sum runs over
all nanoparticle configurations~$s_{\text{nano}}$. In order to avoid
colloidal many-body effects, the infinite dilution limit must be taken
for the microspheres.  The required direct interactions between
nanoparticles and between nanoparticles and microspheres are modeled as
pairwise potentials $V_{\text{nano}}(r)$ and~$V_{\text{m-n}}(r)$,
respectively. Thus, the Hamiltonian can be written as
\begin{equation}
\mathcal{H}(r; \{\mathbf{r}\}) =
V_{\text{micro}}(r) +
\sum_{i,j \in \{\text{nano}\}}  V_{\text{nano}}(|\mathbf{r}_{ij}|)  +
\sum_{i,j \in \{\text{m-n}\}}   V_{\text{m-n}}(|\mathbf{r}_{ij}|) \;,
\label{eq:hamiltonian}
\end{equation}
where the first sum runs over all pairs of nanoparticles, the second sum
runs over all pairs consisting of a microsphere and a nanoparticle, and
$\mathbf{r}_{ij} \equiv \mathbf{r}_j - \mathbf{r}_i$ indicates the pair
separation.  Since $V_{\text{micro}}(r)$ factorizes in
Eq.~(\ref{eq:V_eff}), the nanoparticle-induced pair potential is
independent of the direct microsphere interaction.

In our simulation, both colloidal microspheres and nanoparticles are
modeled as hard spheres with diameters $\sigma_{\rm{micro}}$ and
$\sigma_{\rm{nano}}$, respectively. We approximate the electrostatic
double-layer interactions $V_{\text{nano}}$ and~$V_{\text{m-n}}$ via an
extension of the DLVO theory~\cite{verwey48} to nonidentical particles.
The Debye--H\"uckel approximation is used to linearize the
Poisson--Boltzmann equation. Subsequent application of the Derjaguin
approximation then leads to the Hogg--Healy--Fuerstenau (HHF)
equation~{\cite{hogg66}}. For a pair of particles at surface-to-surface
distance~$D$, under the condition of constant surface potential, this
equation is given by
\begin{eqnarray}\label{eq:ele}
  V_{\rm HHF}(D)
  &=& \frac{1}{2}\epsilon_0\epsilon_r\pi
  \frac{\sigma_{1}\sigma_{2}}{\sigma_{1}+\sigma_{2}}
  ({\Psi_{1}}^2+{\Psi_{2}}^2) \nonumber\\
  &\times& \left\{\frac{2\Psi_{1}\Psi_{2}}{{\Psi_{1}}^2+{\Psi_{2}}^2} 
    \ln\left[\frac{1+\exp(-\kappa D)}{1-\exp(-\kappa D)}\right]
    + \ln[1-{\exp(-2\kappa D)}]\right\} \;,
\end{eqnarray} 
where $\Psi_i$ and $\sigma_i$ ($i=1,2$) represent the surface potential
and diameter, respectively, of a particle of species~$i$, $\epsilon_0$
denotes the vacuum permittivity, $\epsilon_r = 80$ is the dielectric
constant of water, and $\kappa$ is the inverse of the Debye length.

For the interaction between two zirconia nanoparticles, $\Psi_1 = \Psi_2
= \Psi_{\rm{nano}}$ and $\sigma_1 = \sigma_2 = \sigma_{\rm{nano}}$ and
Eq.~(\ref{eq:ele}) reduces to (\cite{hunter01}, Ch.~12)
\begin{equation}
  V_{\text{nano}}(D) = \epsilon_0\epsilon_r\pi \sigma_{\rm nano}\Psi_{\rm
  nano}^2 \ln[1+{\exp(-\kappa D)}] \;.
\label{eq:ele1}
\end{equation}
Due to the strong screening conditions in Ref.~\cite{tohver01}, we can
invoke the approximation $\kappa D \gg 1$ to further simplify this
expression to
\begin{equation}
  V_{\text{nano}}(D) = \epsilon_0\epsilon_r\pi \sigma_{\rm nano}\Psi_{\rm nano}^2
  {\exp(-\kappa D)} \;.
\label{eq:ele1.1}
\end{equation}
While this last approximation overestimates the potential for $\kappa D
\lesssim 1$, i.e., for $D \lesssim \kappa^{-1} = 2{\rm nm}$, we note
that at such small distances even the validity of the original
expression Eq.~(\ref{eq:ele1}) becomes less certain. Furthermore, at
these distances hydration forces become important.  As the hydration
forces have a range of 1--2nm and decay exponentially with a very short
decay length~\cite{israel92}, we account for the hydration shell of the
zirconia particles by replacing $\sigma_{\rm nano}$ with
$\sigma^{'}_{\rm nano} = \sigma_{\rm nano} + 1.5 {\rm nm}$ \emph{solely}
in Eq.~(\ref{eq:ele1.1}).

Since, under the experimental conditions of Ref.~\cite{tohver01}, the
silica microspheres are negligibly charged and thus have
$\Psi_{\rm{micro}} = 0$, we ignore the electrostatic interaction between
silica particles. However, Eq.~(\ref{eq:ele}) shows that an
electrostatic double-layer interaction can arise between the
microspheres and the nanoparticles, even though the former are neutral.
For large size asymmetry $\alpha \equiv \sigma_{\rm micro}/\sigma_{\rm
nano} \gg 1$, Eq.~(\ref{eq:ele}) reduces to
\begin{equation}
  V_{\text{m-n}}(D) = \frac{1}{2}\epsilon_0\epsilon_r\pi \sigma_{\rm
  nano}\Psi_{\rm nano}^2 \ln[1-{\exp(-2\kappa D)}] \;,  
\label{eq:ele2}
\end{equation}
which under the approximation $\kappa D \gg 1$ can be further reduced to
\begin{equation}
  V_{\text{m-n}}(D) = - \frac{1}{2} \epsilon_0\epsilon_r\pi \sigma_{\rm
  nano}\Psi_{\rm nano}^2 {\exp(-2\kappa D)} \;.
  \label{eq:ele2.2}
\end{equation} 

Three important approximations have been made in the derivation of the
HHF equation. First, the linearized Poisson--Boltzmann theory following
from the Debye--H\"uckel approximation is appropriate only for low
surface potentials. For moderate to high surface potentials, $\Psi_i$
$(i=1,2)$ in Eq.~({\ref{eq:ele}}) has to be replaced by an effective
surface potential~$Y_i(D)=4(\Psi_i/y_i^s)\exp(\kappa
D/2)\tanh^{-1}[\exp(-\kappa D/2)\tanh(y_i^s/4)]$ with $y_i^s =
{e\Psi_i}/{k_{\rm B} T}$~\cite{hunter01,sader95}.  As a result, the HHF
equation overestimates the electrostatic interactions for the
experimental value $\Psi_{\rm nano} = 70{\rm mV}$ by up to 25\% at large
separations. For small values of $\kappa D$, the effect is considerably
less (cf.\ Fig.~2 in Ref.~\cite{sader95}).  Secondly, the application of
the Derjaguin approximation makes the HHF equation valid only for small
$\kappa D$.  However, a strict derivation~\cite{sader95} without this
approximation shows that the HHF formula can be rendered applicable to
all $\kappa D$ by replacing
$({\sigma_{1}{\sigma_{2}}})/({\sigma_{1}+\sigma_{2}})$ with
$({\sigma_{1}\sigma_{2}})/({\sigma_{1}+\sigma_{2}+2D})$. For the short
separations and large size ratio investigated here, the difference
between these factors becomes negligible.  Thirdly, a constant surface
potential is assumed in Eq.~(\ref{eq:ele}), although charge-regulating
boundary conditions are more appropriate for the oxide particles
employed in Ref.~\cite{tohver01}.  However, both conditions have been
found to lead to comparable results~\cite{behrens99}, in which an
attraction can arise between neutral and charged
surfaces~\cite{biesheuvel04}. It is important to emphasize that these
mean-field approaches are employed only to set up the direct
colloid--nanoparticle and nanoparticle--nanoparticle \emph{pair
potentials} to be employed in the Monte Carlo simulations.  The actual
simulations explicitly take into account correlation and fluctuation
effects. In addition, as is clear from the use of the HHF equation, the
solvent and all ions are modeled implicitly, yielding a dielectric
medium with a specific electrostatic screening length. While the sheer
number of solvent molecules and ions precludes their explicit
incorporation in the calculation, the strong screening also renders this
largely unnecessary. Another noteworthy point is that the calculations
presented here only yield the effective colloidal pair potential.
Triplet and higher-order interactions, which may become important under
the nondilute conditions in the experiments, are not addressed.

The presence of an attractive interaction between microspheres and
nanoparticles, as found in Eq.~(\ref{eq:ele2.2}), is consistent with the
supernatant measurements reported by Tohver \emph{et
al.}~\cite{tohver01}. Also their zeta-potential
measurements~\cite{tohver01} (which exploit the electrophoretic mobility
of the microspheres) indicate a certain degree of
microsphere--nanoparticle association, since the increase of the zeta
potential with increasing nanoparticle concentration requires a
cooperative motion of colloids and nanoparticles.  Accordingly, it might
be tempting to dismiss the haloing phenomenon as a a standard case of
electrostatic stabilization resulting from an effective charge build-up
on the colloids. However, as is shown below, the situation is
considerably more complicated, since the resulting halo is dynamic in
nature and leads to effective interactions that are attractive,
repulsive, or oscillatory, depending on nanoparticle concentration.

\section{Simulation details}

As already noted in Ref.~\cite{geomc}, traditional computational
approaches experience severe difficulties in the treatment of this
system.  Inherent limitations in most computational methodologies
greatly restrict the range of accessible size ratios in multi-component
mixtures because larger species tend to get trapped by the smaller.
This ergodicity problem becomes more severe if small particles
accumulate around larger ones, as is the case in the present
system~\cite{tohver01,tohver01b}. The simulation methods that are
capable of explicitly simulating both large and small species are
limited to size ratios up to $\alpha = 10$~\cite{malherbe99}, except for
a few cases of hard-sphere mixtures.  Thus, an alternative route that is
sometimes taken is to only simulate the smaller species around a pair of
immobile large particles at a fixed
separation~\cite{dickman97,allahyarov98,wu98}.  Since the effective
force can be evaluated only for one separation at a time, a series of
such simulations for different large-particle distances must be carried
out to construct the entire curve for the effective force.  Since the
large-particle concentration must be kept very small to exclude
many-body effects, the total particle number typically becomes very
large if the small species is present at a fixed concentration and
$\alpha$ is large. As a result, this method becomes impractically slow
even for moderate accuracy.

We overcome this problem by exploiting a novel, highly efficient Monte
Carlo scheme~\cite{geomc,liu05a}, which permits the explicit inclusion
of interacting species of vastly different sizes. This algorithm is a
variant of the geometric cluster algorithm originally introduced by
Dress and Krauth~\cite{dress95}. The explicit simulation of both species
makes it possible to calculate the effective interactions directly from
the inversion of the particle pair correlation function $g(r)$ with less
computational effort than previous techniques. In the infinite dilution
limit, the potential of mean force follows from $V_{\rm {eff}}(r) =
-k_{\rm B}T \ln g(r)$. We employ only two microspheres in a periodic
cubic box at concentration $\phi_{\rm {micro}} = 0.002$. In principle,
the potential of mean force must be calculated from simulations in which
the nanoparticles are kept at constant chemical potential. For
efficiency reasons, we instead use the canonical ensemble, in which the
number of nanoparticles is constant. Since our simulation cell is very
large, fluctuations in the number of nanoparticles are expected to be
small as the microsphere separation varies, and consequently the
difference between the two ensembles is expected to be very
small~\cite{wu98}. In addition, we have explicitly verified that no
changes occur in the effective potential as the box size is increased
further (i.e., $\phi_{\rm {micro}}$ is lowered further).

\begin{table}
  \caption{Computational parameters for the various nanoparticle
  concentrations, $\phi_{\rm nano}$, investigated. All data apply to a
  nanoparticle--microsphere mixture containing two microspheres at a
  volume fraction $\phi_{\rm micro}=0.002$, with a
  colloid--nanoparticle size ratio~$\alpha=100$. $N_{\rm nano}$
  indicates the number of nanoparticles in the simulation cell.
  $\smash{\langle N^{\rm cluster}_{\rm micro}\rangle}$ and
  $\smash{\langle N^{\rm cluster}_{\rm nano}\rangle}$ indicate the
  average number of microspheres and nanoparticles, respectively, in
  each cluster. Note that each cluster is started from a
  microsphere. Also shown are the total number of clusters constructed
  and the required CPU time.}
  \label{tab:table1}
\renewcommand{\tabcolsep}{4mm}
\renewcommand{\arraystretch}{1.5}
\begin{tabular}{|c|r|c|r|r|c|}
\hline
\multicolumn{1}{|c|}{$\phi_{\rm nano}$}   &
\multicolumn{1}{c|}{$N_{\rm nano}$}       & 
\multicolumn{1}{c|}{$\langle N^{\rm cluster}_{\rm micro}\rangle$} &
\multicolumn{1}{c|}{$\langle N^{\rm cluster}_{\rm nano\vphantom{k}}\rangle$}  &
\multicolumn{1}{c|}{No.\ of Clusters} &
\multicolumn{1}{c|}{CPU hours}
\\ \hline
  0.0007      &   700,000 & 1.0089 &  2503  & 200,000,000  & 1,000 \\ \hline
  0.0010      & 1,000,000 & 1.0089 &  2626  & 200,000,000  & 1,800 \\ \hline
  0.0030      & 3,000,000 & 1.0092 &  5791  &  50,000,000  & 1,500 \\ \hline
  0.0050      & 5,000,000 & 1.0098 &  8737  &  40,000,000  & 3,100 \\ \hline
  0.0070      & 7,000,000 & 1.0110 & 11808  &  60,000,000  & 4,500 \\ \hline 
\end{tabular}
\end{table}

In the actual calculations of the potential of mean force, $\phi_{\rm
{nano}}$ is varied from $0.0007$ to $0.01$, which encompasses the entire
range over which stabilization is observed in the experiments. Following
the experimental conditions~\cite{tohver01,tohver01b}, we choose a
microsphere diameter $\sigma_{\rm micro} = 0.6\mu{\rm m}$ and a
nanoparticle diameter $\sigma_{\rm nano}=6{\rm nm}$. Furthermore, we set
$\Psi_{\rm nano} = 70 \rm mV$ and $\kappa \sigma_{\rm nano} = 3.0$,
except when explicitly indicated otherwise. Since the
colloid--nanoparticle size ratio is $\alpha = 100$, the resulting number
of nanoparticles reaches $N_{\rm nano}=7 \times 10^6$ for $\phi_{\rm
nano}=0.007$ (for $\phi_{\rm nano}=0.01$, where $N_{\rm nano} = 10^7$,
our calculations are only accurate enough to locate the minimum in the
effective potential). The corresponding computational effort is
summarized in Table~\ref{tab:table1}.

A crucial ingredient to accelerate the simulations is the use of the
cell index method with linked lists. Here, two independent sets of
linked cells are employed, with cell sizes based upon the respective
interaction cutoffs used.  In addition, a direct mapping is used to
compute the potential $V_{\text{m-n}}$ between species that reside in
different cell sets~\cite{liu05a}.

\section{Results and discussion}

\subsection{Nanoparticle adsorption}

In order to investigate the validity of the interactions derived in
Sec.~\ref{sec:interaction}, we first carry out Monte Carlo
simulations using the approximate expressions in Eqs.\ (\ref{eq:ele1.1})
and~(\ref{eq:ele2.2}). The interactions are cut off at $r_c^{\rm
nano}=5\sigma_{\rm nano}$ for $V_{\text{nano}}(r)$ and $r_c^{\rm
m-n}=60\sigma_{\rm nano}$ for $V_{\text{m-n}}(r)$.  Unlike in the
calculations of the potential of mean force (presented below), here we
wish to explicitly account for colloidal many-body effects and thus use
a system containing 50 microspheres at the same volume fraction as the
experiments, $\phi_{\rm micro}=0.1$. Up to $5 \times 10^6$ nanoparticles
are present.

\begin{figure}
  \centering\includegraphics[width=\widefigurewidth]{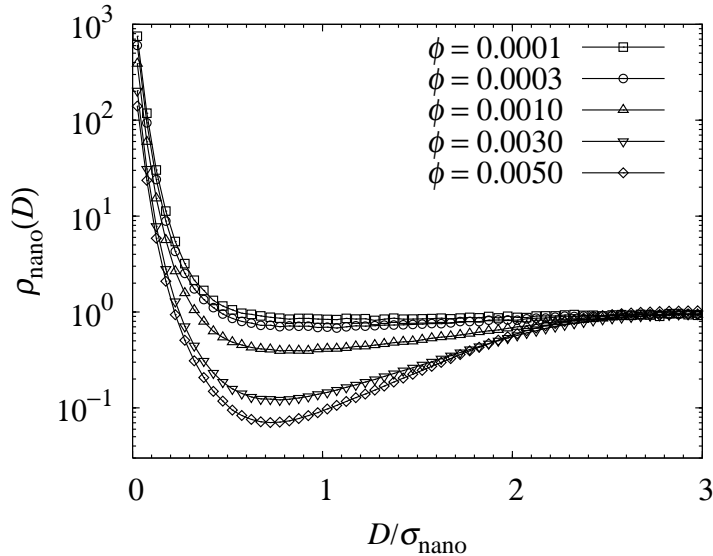}
  \caption{Log--linear plot of the normalized density profile $\rho_{\rm
  nano}(D)/\rho_{\rm nano}(\infty)$ of nanoparticles around a
  microsphere as a function of the colloid--nanoparticle surface
  separation~$D$, for various nanoparticle concentrations $\phi_{\rm
  {nano}}$. The accumulation near contact ($D=0$) results from the
  attractive potential. The minimum that develops with
  increasing~$\phi_{\rm nano}$ is used to define the extent of the
  halo.}
  \label{fig:ads1}
\end{figure}

\begin{figure}
  \centering \includegraphics[width=\widefigurewidth]{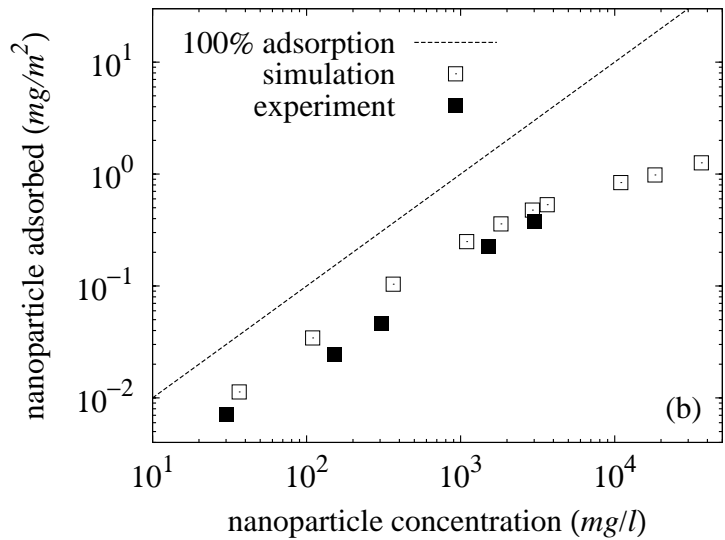}
  \caption{Nanoparticle adsorption per colloidal microsphere as a
  function of nanoparticle concentration, at fixed microsphere
  concentration, $\phi_{\rm micro}=0.1$. The agreement between experiment
  (solid squares, Ref.~\cite{tohver01}) and simulation data (open
  squares) is quite reasonable, in view of the approximate potentials
  employed. The highest experimental concentration, $3650{\rm mg/l}$,
  corresponds to $\phi_{\rm nano} = 10^{-3}$.}
  \label{fig:ads2}
\end{figure}

As illustrated in Fig.~\ref{fig:ads1}, the nanoparticle
distribution~$\rho_{\rm nano}(D)$ around a microsphere (expressed as a
function of the microsphere--nanoparticle surface separation~$D$) has a
very strong peak near contact, indicating the accumulation of
nanoparticles around the colloids.  The thickness of this accumulation
layer, as determined by the minimum of $\rho_{\rm nano}(D)$, is
approximately $\sigma_{\rm nano}$.  For low volume fractions, there is
only a negligible difference between the density profiles for
different~$\phi_{\rm nano}$.  However, for higher volume fractions,
$\phi_{\rm nano} \ge 0.001$, the nanoparticle distribution starts to
vary, with a more rapid decrease with increasing~$D$ and a clear
minimum.  By integrating $\rho_{\rm nano}(D)$ from contact to this
minimum, we compute the amount of adsorbed nanoparticles, shown in
Fig.~\ref{fig:ads2} along with the experimental
findings~\cite{tohver01}.  As can be seen, the agreement with the
experiments is fairly good. Both data sets exhibit a linear dependence
on $\phi_{\rm nano}$ and a degree of adsorption that is far below 100\%.
The systematic overestimation in the simulations is compatible with the
high nanoparticle surface potential $\Psi_{\rm nano} = 70{\rm mV}$,
which lies somewhat above the regime of validity of the Debye--H\"uckel
approximation employed in the HHF equation.  At volume fractions above
$10^{-3}$ the adsorption saturates owing to mutual repulsion between the
charged nanoparticles. We conclude that the pair potentials derived in
Sec.~\ref{sec:interaction} constitute a representative description of
the experimental system and capture the essential interactions.

\subsection{Effective colloidal interactions and colloidal stabilization}

\begin{figure}
  \centering 
  \includegraphics[width=\figurewidth,angle=0]{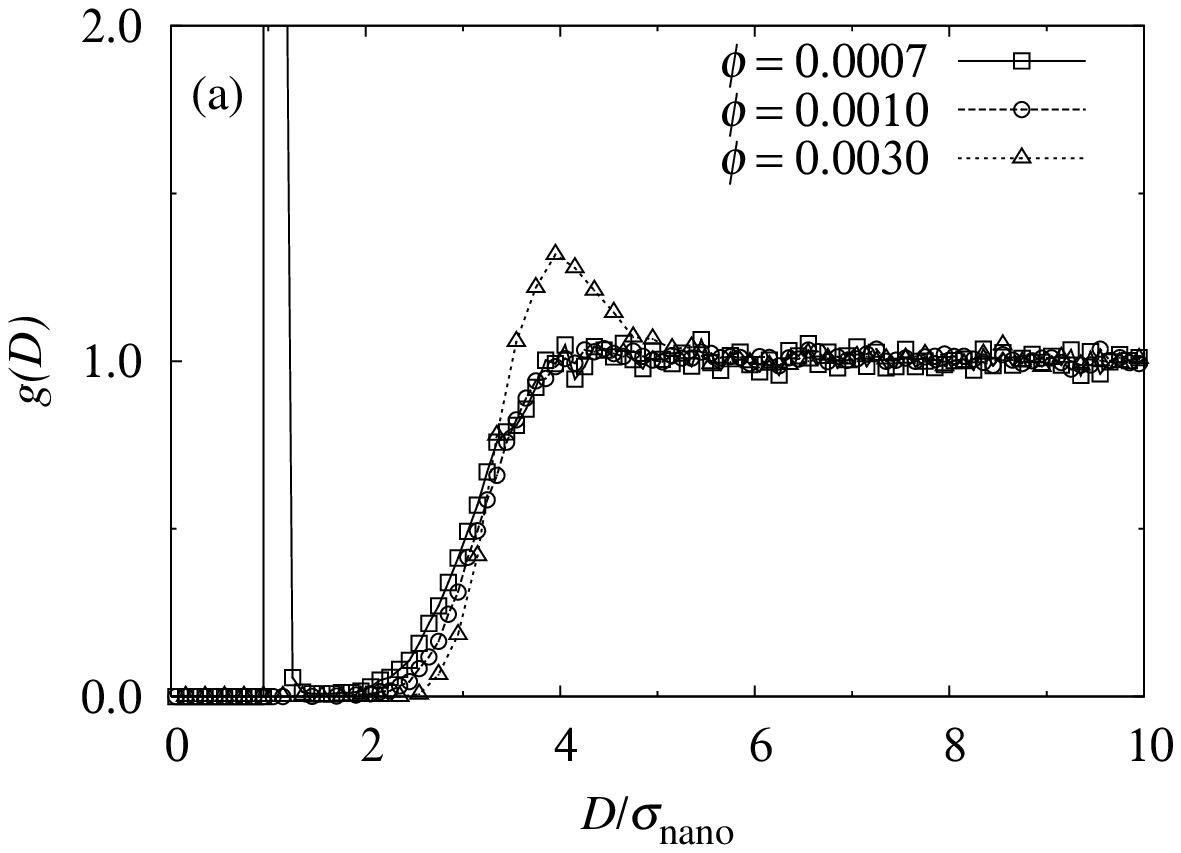}
  \includegraphics[width=\figurewidth,angle=0]{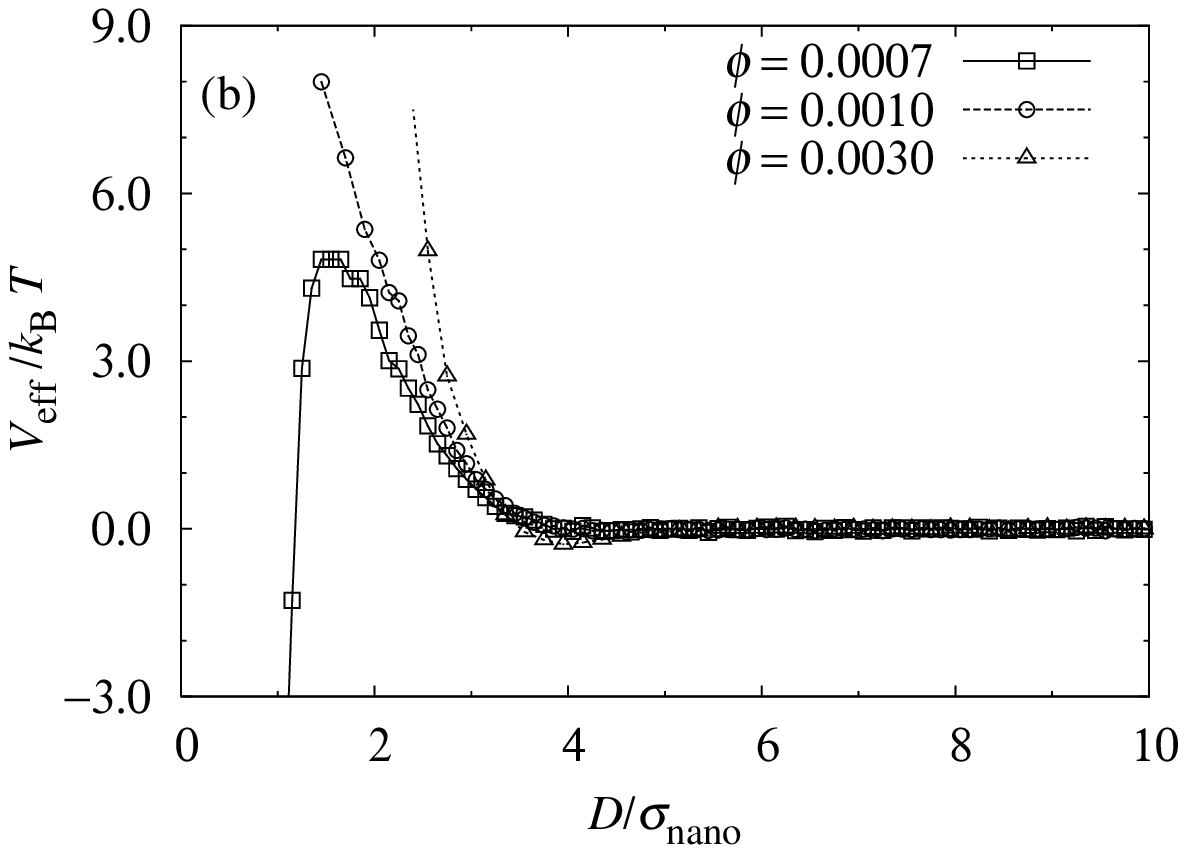}
  \caption{(a) Radial microsphere distribution function $g(D)$ as a
  function of surface-to-surface separation~$D$, for nanoparticle
  concentrations that correspond to the stable fluid regime. (b)
  Corresponding effective potential $V_{\rm eff}/k_{\rm B}T$ between a
  pair of colloidal microspheres. At these low volume fractions, the
  effective interaction shows a repulsive barrier that prevents
  colloidal aggregation.}
  \label{fig:low}
\end{figure}

Since the direct interactions have been shown to provide a fairly
accurate representation of the experimental system, we now proceed to
calculate the resulting \emph{effective} pair potential between the
colloids, as a function of nanoparticle volume fraction~$\phi_{\rm
nano}$.  At very low $\phi_{\rm nano} = 7 \times 10^{-4}$ the
microsphere pair correlation function [Fig.~\ref{fig:low}(a)] shows a
very strong peak at a surface-to-surface distance $D=\sigma_{\rm nano}$,
corresponding to configurations in which the pair of microspheres is
separated by a single nanoparticle. However, this peak is followed by a
rather broad ``exclusion zone,'' corresponding to pair separations that
are extremely unlikely to occur. While the short-distance (``bridging'')
peak is only observed for $\phi_{\rm nano}= 7 \times 10^{-4}$, we expect
it to be present also for the other concentrations in
Fig.~\ref{fig:low}(a), but the effective barrier between $D=\sigma_{\rm
nano}$ and $D \approx 2\sigma_{\rm nano}$ makes it very unlikely for the
system to arrive in such a configuration during the course of the
simulation.  Indeed, the potential of mean force, derived from the
microsphere pair correlation function and shown in
Fig.~\ref{fig:low}(b), displays a repulsive barrier of $5k_{\rm B}T$ at
$\phi_{\rm nano} = 7 \times 10^{-4}$. If $\phi_{\rm nano}$ is increased
to $1.0 \times 10^{-3}$ and $3.0 \times 10^{-3}$, the height of this
barrier increases even further.

Since the van der Waals attraction between microspheres is additive to
the nanoparticle-induced interaction (cf.\ Sec.~\ref{sec:interaction}),
it is omitted in the effective potential shown in Fig.~\ref{fig:low}(b).
While it is strongly attractive at short separations, it decays rapidly
with increasing microsphere separation and has a strength of
approximately only $-1k_{\rm B}T$ at a surface-to-surface separation
$D=\sigma_{\rm nano}$~\cite{tohver01b}. As a result, the repulsive
barrier persists in the \emph{net} pair interaction and is, for
$\phi_{\rm nano} \gtrsim 7 \times 10^{-4}$, sufficient to prevent
gelation, resulting in kinetic stabilization of the suspension.  The
corresponding threshold agrees remarkably well with the experimentally
observed gel--fluid transition near $\phi_{\rm nano} = 5 \times
10^{-4}$~\cite{tohver01}.

\begin{figure}
  \centering \includegraphics[width=\figurewidth,angle=0]{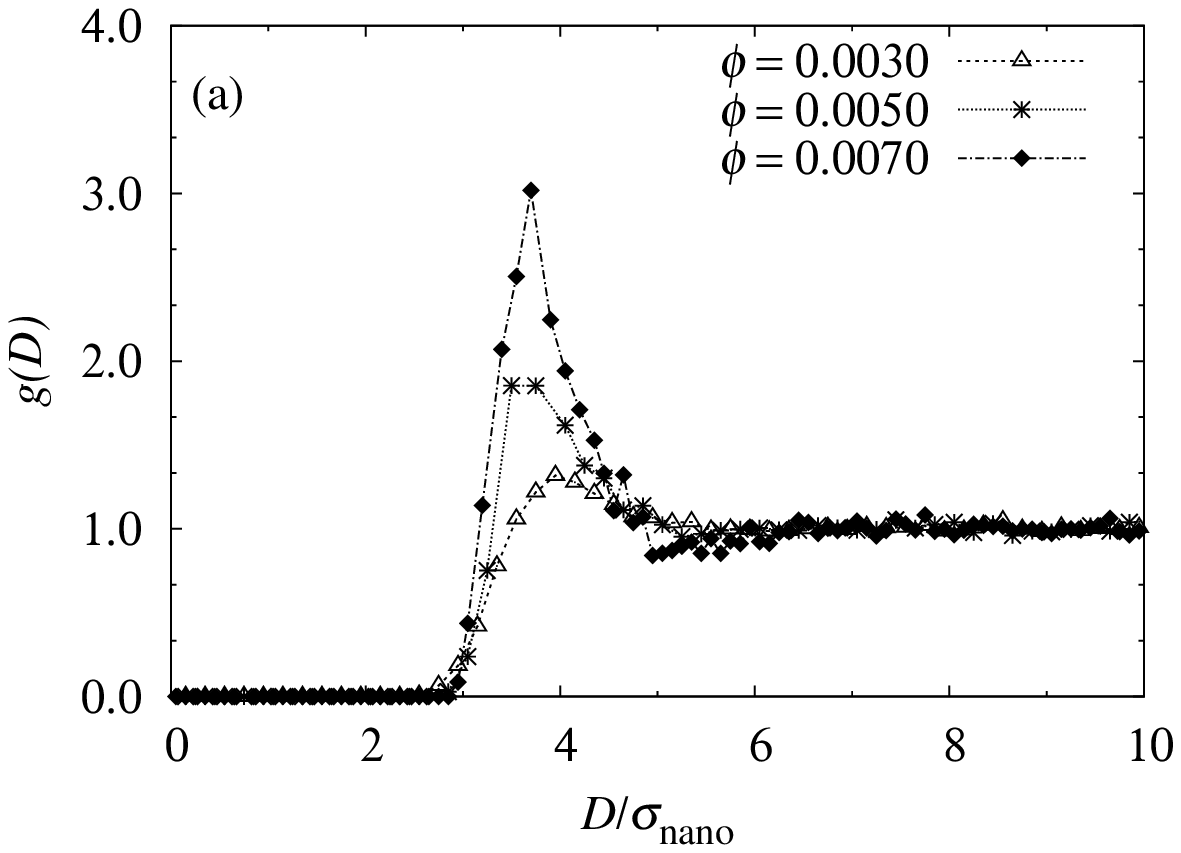}
  \includegraphics[width=\figurewidth,angle=0]{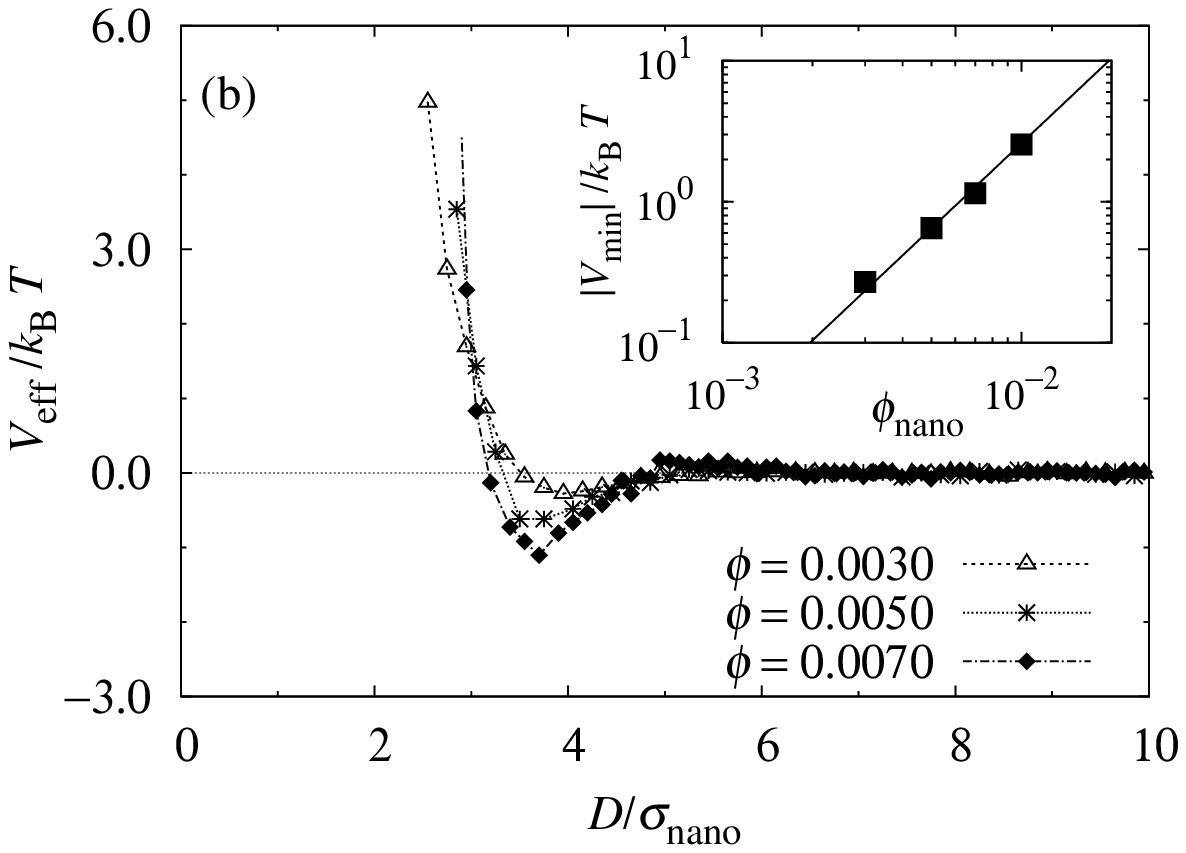}
  \caption{Counterpart of Fig.~\ref{fig:low}, but now for higher
  nanoparticle volume fractions. A peak develops in the radial
  distribution function (a) for surface separations near $4\sigma_{\rm
  nano}$. The corresponding attractive minimum in the effective
  potential (b) grows quadratically with $\phi_{\rm nano}$, as shown in
  the inset. This attraction is responsible for reentrant gelation of
  the colloidal microspheres.}
  \label{fig:high}
\end{figure}

As foreshadowed by the local maximum in $g(D)$ for $D \approx 4
\sigma_{\rm nano}$ at $\phi_{\rm nano}=3.0 \times 10^{-3}$
[Fig.~\ref{fig:low}(a)], the situation changes if the nanoparticle volume
fraction is raised further. The peak shifts to slightly lower
separations and, more importantly, increases in height, see
Fig.~\ref{fig:high}(a). This corresponds to the emergence of an
effective attraction, which reaches a strength of nearly $-3 k_{\rm B}T$
at $\phi_{\rm nano} = 10^{-2}$ [Fig.~\ref{fig:high}(b)]. Indeed, since
this strength corresponds to the typical estimate for the onset of
colloidal gelation, we identify $\phi_{\rm nano} = 10^{-2}$ as the
approximate upper bound of the stable fluid region. Again, this is in
quite good agreement with the experimentally observed reentrant gelation
at $\phi_{\rm nano} \approx 5 \times 10^{-3}$, especially in view of the
approximate interactions potentials employed.

A natural explanation for the reentrant behavior might be found in a
depletion interaction~\cite{asakura54} induced by the nanoparticles, as
suggested in Ref.~\cite{tohver01}. However, as shown in the inset of
Fig.~\ref{fig:high}(b), the attractive strength increases quadratically
with $\phi_{\rm nano}$, whereas the classical AO theory for
depletion~\cite{asakura54} predicts an attractive minimum at contact
($D=0$), with a strength that has a linear concentration dependence.
Although the quadratic dependence is observed for only less than a
decade in $\phi_{\rm nano}$, and calculations for larger nanoparticle
concentrations are computationally prohibitively expensive, the
nonlinearity of the data appears quite pronounced.  Since the attractive
minimum in Fig.~\ref{fig:high}(b) does not occur at contact, one could
envisage a situation in which both colloids are decorated with a layer
of nanoparticles (effectively increasing the colloid diameter) and the
remaining nanoparticles induce a depletion interaction. While this might
account for the position of the potential minimum, it does not account
for the quadratic concentration dependence, which persists even after
the corresponding adjustment has been made to the nanoparticle
concentration employed in the inset of Fig.~\ref{fig:high}(b). An
alternative explanation for the deviation from the linear concentration
dependence might be that the strong mutual electrostatic repulsion leads
to \emph{effective} nanoparticle concentrations that considerably exceed
the bare concentrations, causing the system to leave the regime of
validity of the AO theory.  However, even the effective concentrations
are so low that no significant deviations from linearity are expected.
Thus, we conclude that the effective attractions do not result from a
classical depletion mechanism alone. Given the nonadditive nature of the
interactions, this is not truly surprising. To support this, an
examination of relevant particle configurations shows no significant
nanoparticle depletion between a pair of microspheres at a separation $D
\approx 4\sigma_{\rm nano}$. It appears therefore plausible that the
microsphere attraction arises from correlations between the highly
charged nanoparticles in the gap between the colloids.

\begin{figure}
  \centering
  \includegraphics[width=\figurewidth,angle=0]{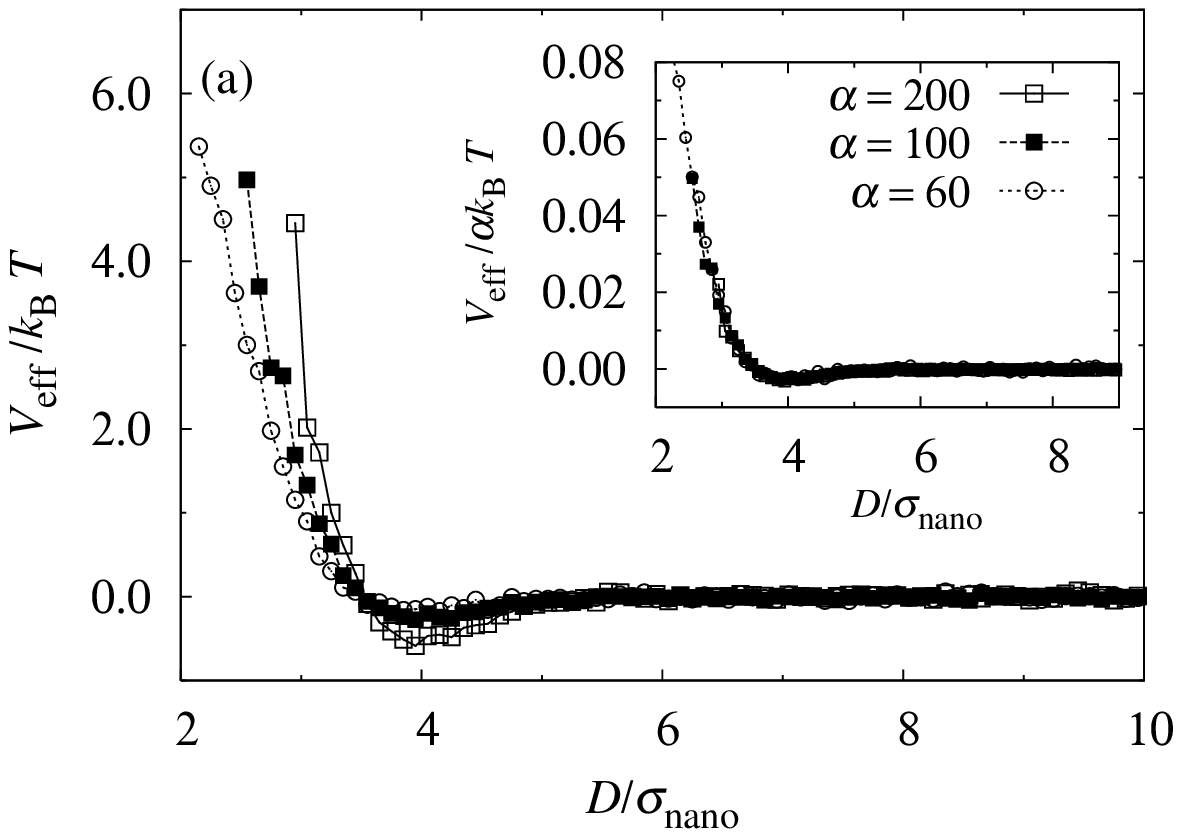}
  \includegraphics[width=\figurewidth,angle=0]{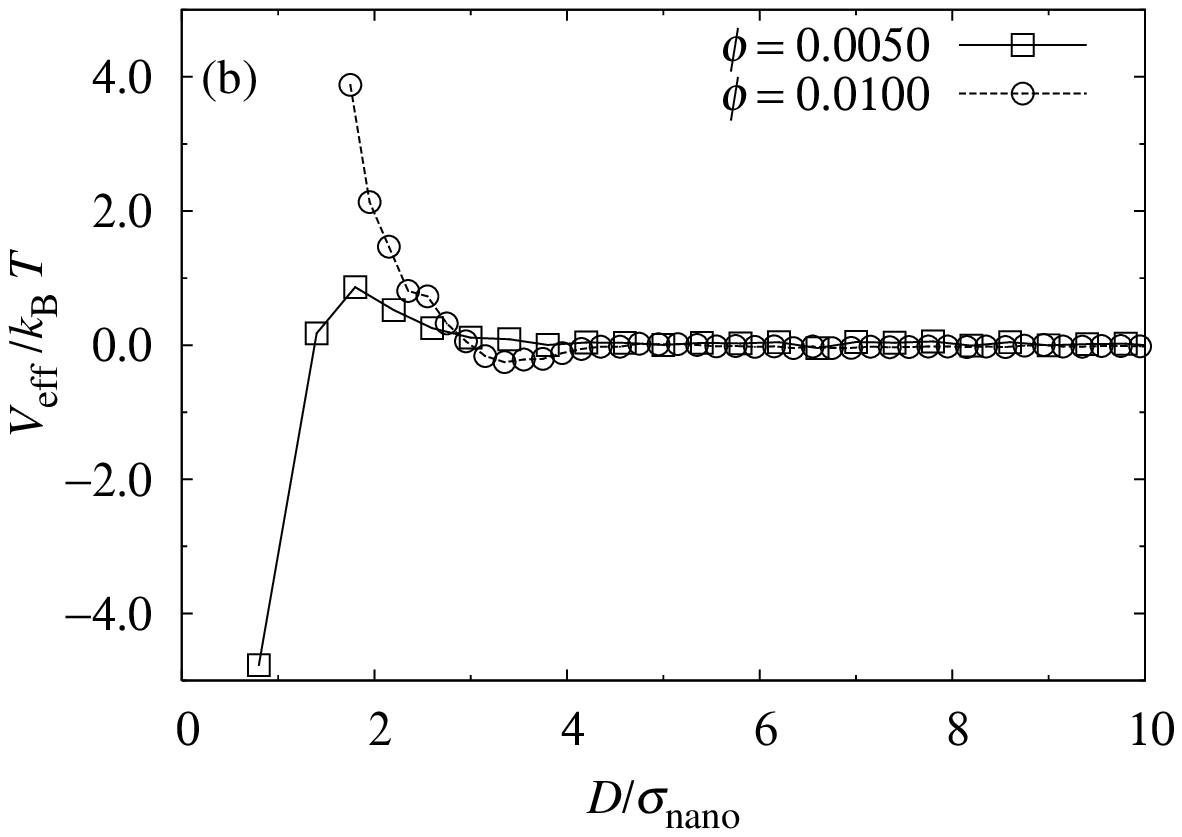}
  \caption{(a) Effect of nanoparticle--colloid size asymmetry~$\alpha$
  on the effective colloidal interactions. At fixed nanoparticle size
  and volume fraction $\phi_{\rm nano}=0.003$, both the strength of the
  repulsion and the depth of the attractive minimum increase with
  colloid size. As shown in the inset, the effective interactions show a
  perfectly linear scaling with $\alpha$, indicating that the Derjaguin
  approximation applies (see text).  (b) Effect of nanoparticle charge.
  All parameters are identical to those in Fig.~\protect\ref{fig:low},
  except for the nanoparticle surface potential, which has been lowered
  from $70{\rm mV}$ to $50{\rm mV}$.  As a result, much higher
  nanoparticle volume fractions are required to achieve an appreciable
  colloidal repulsion.}
  \label{fig:parameters}
\end{figure}

Since an understanding of the role of size and charge asymmetry between
colloids and nanoparticles is of crucial importance for potential
applications, we have investigated several other parameter combinations.
Employing numerical data from Ref.~\cite{martinez05}, we have been able
to extend the size asymmetry to $\alpha=200$ (which involved a system
with $12 \times 10^6$ nanoparticles).  Figure~\ref{fig:parameters}(a)
displays the effective interaction resulting from a fixed nanoparticle
volume fraction, $\phi_{\rm nano}=0.003$, for colloids of diameter
$0.36\mu{\rm m}$, $0.60\mu{\rm m}$, and $1.20\mu{\rm m}$ (size ratio
$\alpha = 60$, $100$, and $200$, respectively). For larger $\sigma_{\rm
micro}$ both the height of the repulsive barrier and the depth of the
attractive minimum in the effective interaction are enhanced.  This is
consistent with the experimental observation~\cite{tohver01} that both
boundaries of the stable fluid region were lowered when the size of the
colloidal microspheres was increased. The inset of
Fig.~\ref{fig:parameters}(a) shows the same potentials scaled by the
size asymmetry. The excellent overlap of all curves shows that the
dependence on size asymmetry is almost perfectly linear. Thus, the
effective interactions follow the Derjaguin approximation, according to
which the force between two large, identical spheres of
diameter~$\sigma$ can be expressed in terms of the interaction energy of
two parallel plates $W(D)$~\cite{israel92},
\begin{equation}
F^{\rm Derjaguin}(D) \approx \frac{\pi\sigma}{2} \int_{D}^{\infty} f(z)dz =
\frac{\pi\sigma}{2} W(D) \;,
\label{eq:Derjaguin}
\end{equation}
where $f(z)$ is the normal force per unit area between two flat
surfaces. This relation, which is applicable to any type of force law as
long as the separation $D$ is much less than the sphere
diameter~$\sigma$, indeed predicts that the colloidal force (and hence
the interaction potential) has a linear dependence on~$\sigma$, as an
increase of the colloidal diameter does not affect $W(D)$. We note that
for actual colloidal stabilization not only the size ratio, but also the
\emph{absolute size} of the nanoparticles plays a role, as it determines
the range of the repulsive barrier (which must exceed the range of the
van der Waals interaction).

Variation of the nanoparticle charge has multiple effects, since it not
only alters the nanoparticle repulsion (and hence their effective volume
fraction and the halo--halo interaction), but also the formation of the
halo itself through the colloid--nanoparticle attraction,
Eq.~(\ref{eq:ele2.2}). This is also seen by expressing the effective
force between a pair colloids at surface separation~$D$ in terms of the
nanoparticle distribution~$\rho(\mathbf{r'}; D)$, where $\mathbf{r'}$ is
measured with respect to the first colloid,
\begin{equation}
\mathbf{F}(D) = - \int \rho(\mathbf{r'};D)
\frac{\partial V_{\text{m-n}}(|\mathbf{r'}|)}{\partial \mathbf{r'}}
\, d\mathbf{r'} \;.
\label{eq:eff_force}
\end{equation}
The nanoparticle charge will affect both $\rho(\mathbf{r'}; D)$ and
$V_{\text{m-n}}(|\mathbf{r'}|)$. Since a systematic variation of the
nanoparticle charge has not been pursued in experiments, we present
simulation results for the zirconia--silica system with $\alpha=100$,
but at a nanoparticle surface potential that has been lowered from
$\Psi_{\rm nano}=70{\rm mV}$ to $\Psi_{\rm nano} = 50{\rm mV}$, see
Fig.~\ref{fig:parameters}(b).  The effective interactions exhibit the
same trend as for higher surface potentials, but at considerably higher
volume fractions.  Indeed, at $\Psi_{\rm nano} = 50{\rm mV}$ a given
effective colloidal repulsion is reached only at a volume fraction that
is approximately ten times larger than for $\Psi_{\rm nano} = 70{\rm
mV}$.

Finally, we have performed control simulations in which no attraction is
present between the nanoparticles and the microspheres. Under these
conditions, neither the observed nanoparticle adsorption
(Fig.~\ref{fig:ads2}) nor the effective repulsions (Fig.~\ref{fig:low})
are recovered, except at nanoparticle concentrations that are
\emph{much} higher than in the experiment or if the nanoparticle
repulsion has a much longer range than implied by the actual screening
length $\kappa^{-1} \approx \sigma_{\rm nano}/3$~\cite{tohver01b}.

\subsection{Nanoparticle distributions}

Since the effective potentials essentially result from the density
distribution of the nanoparticles [cf.\ Eq.~(\ref{eq:eff_force})], we
have examined this distribution $\rho(\mathbf{r'};D)$ directly for a
pair of colloids at fixed separation~$D$, in order to gain insight in
the origin of these potentials. Exploiting the symmetry around the
$z$-axis passing through the centers of both colloids, we express the
nanoparticle positions in cylindrical coordinates $(r,\phi,z)$. A
rotational average is performed and the $r$ and $z$ coordinates of the
nanoparticles are binned. For each bin, the local nanoparticle density
is computed. The data shown here apply to a smaller size ratio
$\alpha=40$ (i.e., $\sigma_{\rm micro}=0.24\mu{\rm m}$), merely for
reasons of clarity.  The nanoparticle concentration is set to $\phi_{\rm
nano}=0.003$, resulting in a strongly repulsive effective interaction.
Since the microspheres are fixed during the simulation, the conventional
Metropolis Monte Carlo method is used to simulate the movement of the
nanoparticles. The averages are obtained over 5,000 independent
samples.

\begin{figure}
  \centering
  \includegraphics[width=0.35\textwidth,angle=0]{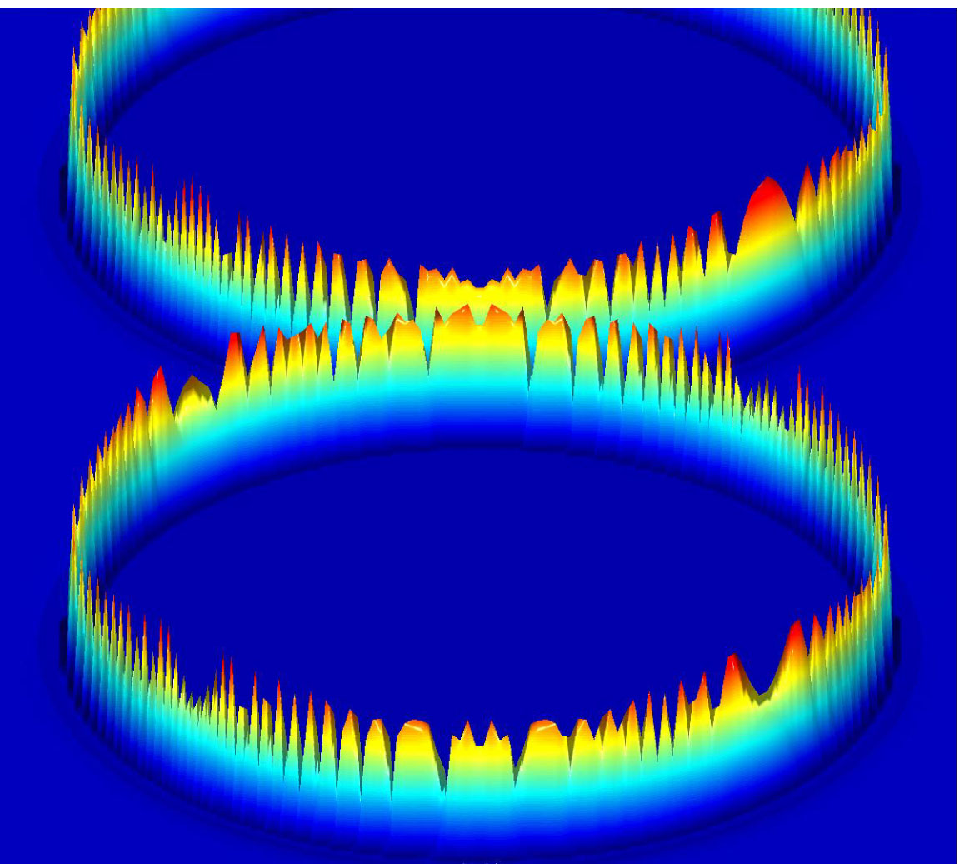}
  \includegraphics[width=0.35\textwidth,angle=0]{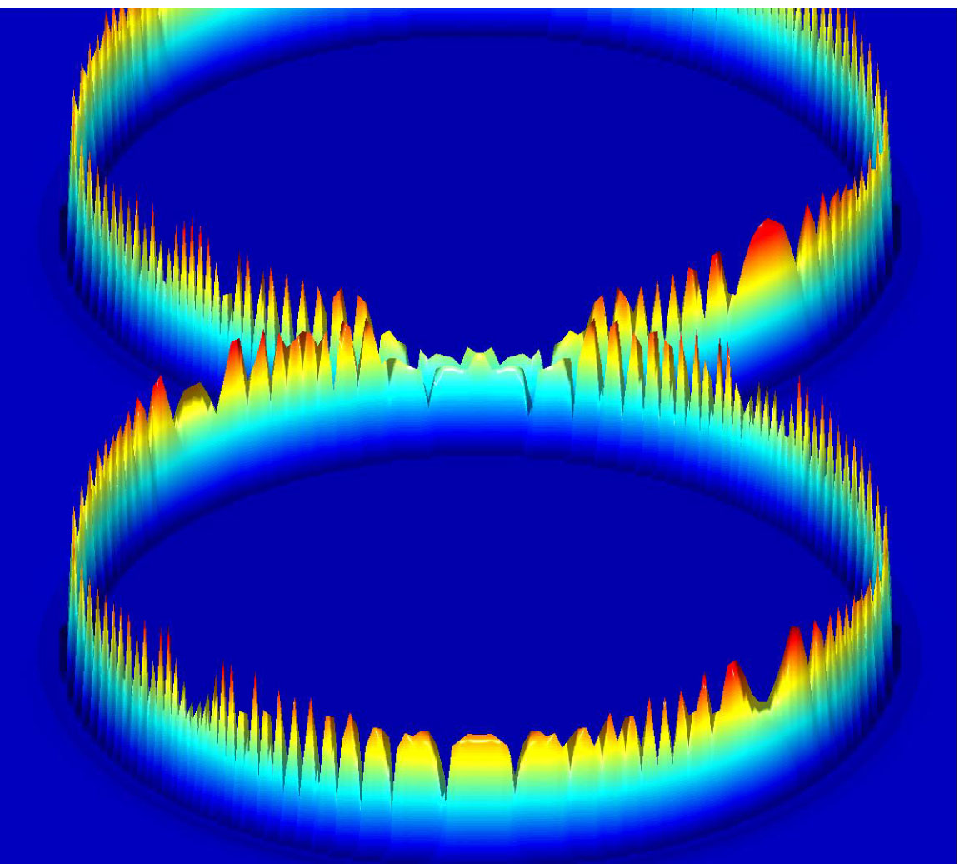}
  \caption{[color online] Nanoparticle density distribution around a
  pair of colloidal particles at a surface-to-surface distance~$D$. (a)
  At larger separations ($D=5.0\sigma_{\rm nano}$), the halos around
  each colloid are uniform and no appreciable effective interaction
  occurs. (b) Upon closer approach ($D=1.8\sigma_{\rm nano}$), a
  \emph{depletion} of nanoparticles in the gap region is observed, but
  the effective colloidal interaction is strongly \emph{repulsive},
  indicating that energetic terms dominate over entropic terms in the
  effective interaction. The jagged appearance of the distributions
  results from the binning procedure.}
  \label{fig:distr}
\end{figure}

At a surface separation $D = 5.0\sigma_{\rm nano}$
[Fig.~\ref{fig:distr}(a)] nanoparticles accumulate in the vicinity of
the colloidal surface and form a uniform, thin halo around each colloid.
In this case, the effective interaction is negligibly small. If the
colloids are forced closer together, the mutual repulsion between the
nanoparticles leads to a redistribution of particles in the halo. At a
separation $D=1.8\sigma_{\rm nano}$ [Fig.~\ref{fig:distr}(b)] most
nanoparticles have been expelled from the gap region between the
colloids and a depletion zone appears. While reminiscent of the
depletion effect in hard-sphere mixtures~\cite{asakura54}, the effective
colloidal interaction in this case is strongly \emph{repulsive} rather
than attractive. This indicates that the energy penalty resulting from
nanoparticle rearrangements dominates over depletion-like entropic
effects.  As discussed in the context of Fig.~\ref{fig:high}(b), at a
separation where a weak effective attraction is observed, namely at $D
\approx 4.0\sigma_{\rm nano}$, the nanoparticle distribution looks very
similar to that depicted in Fig.~\ref{fig:distr}(a) for $D =
5.0\sigma_{\rm nano}$, i.e., no depletion is observed, and detection of
significant differences associated with the observed attraction probably
requires a spatial resolution (bin size) and statistical accuracy that
exceeds what is obtained in the current work.

\section {Comparison to integral-equation theory}

The system studied in this paper has also been investigated by means of
integral-equation theory~\cite{chavez04,karanikas04}, in which the
effective colloidal pair potential~$V_{\text{micro}}^{\rm eff}$ is
computed using the two-component Ornstein--Zernike (OZ) equations
supplemented with the hypernetted-chain (HNC) closure. Ch\'avez \emph{et
al.}~\cite{chavez04} were the first to demonstrate that, under the
appropriate conditions, the electrostatic repulsion between
nanoparticles alone is sufficient to cause the formation of a
nanoparticle monolayer on the microsphere surface. Depending on size
asymmetry and nanoparticle concentration, it is also possible that the
nanoparticles are less strongly bound and that a diffuse halo emerges
instead of a monolayer. Rather than aiming to precisely reproduce the
experimental conditions, Ref.~\cite{chavez04} focuses on general trends.
In particular, it is emphasized that, as a general criterion for the
formation of a halo or monolayer, the microsphere diameter must be
compared to the average (bulk) nanoparticle separation. Evidently, this
criterion relies on a sufficiently strong nanoparticle repulsion.
Indeed, all calculations were performed for a specific (large) choice
for the screening length, $\kappa \sigma_{\rm nano}=0.15$, i.e.,
$\lambda \approx 6.67\sigma_{\rm nano}$. If seems doubtful that this
criterion is applicable for the very different conditions of
Ref.~\cite{tohver01}.

Subsequently, and coincident with the simulation results~\cite{liu04b},
another integral-equation study was performed by Karanikas and
Louis~\cite{karanikas04}. They find that, for a suitable parameter
choice, the effective microsphere interaction exhibits a dependence on
nanoparticle concentration~$\phi_{\rm nano}$ that is qualitatively
compatible with the experimental findings.  Initially, and just as found
already by Ch\'avez \emph{et al.}~\cite{chavez04}, the nanoparticles
induce a repulsive interaction that increases in strength with
nanoparticle concentration, resulting in colloidal stabilization at a
concentration~$\phi_{\rm nano}^{\rm lower}$.  Upon further increase of
$\phi_{\rm nano}$, an attractive minimum appears in
$V_{\text{micro}}^{\rm eff}$, which becomes sufficiently strong to lead
to reentrant gelation of the colloids at a concentration~$\phi_{\rm
nano}^{\rm upper}$.

While the general trends observed in Ref.~\cite{karanikas04} are
certainly encouraging, we note that there are also important
quantitative discrepancies with the experimental~\cite{tohver01} and
simulational~\cite{liu04b} findings. Various calculations in
Ref.~\cite{karanikas04} employ a Debye screening length of $\lambda =
5.0 \sigma_{\rm nano}$, much larger than the experimental value~$\lambda
\approx 0.33\sigma_{\rm nano}$ and, incidentally, rather close to the
choice of Ch\'avez \emph{et al.}~\cite{chavez04}.  This increases the
mutual repulsion between nanoparticles, to the extent where it may alter
the mechanism responsible for the observed stabilization. Conversely, in
Figure~3 of Ref.~\cite{karanikas04} it is shown that for parameter
choices closer to the experimental conditions (e.g., point ``A'' in this
figure, corresponding to a screening length $\lambda = \sigma_{\rm
nano}$ and a nanoparticle contact energy $\varepsilon_{\rm
nano}=6.0k_{\rm B}T$), the lower bound of the stability window is
shifted to nanoparticle volume fractions that are one to two orders of
magnitude larger than in Refs.~\cite{tohver01,liu04b}, whereas the
window is also considerably too narrow, $\phi_{\rm nano}^{\rm
upper}/\phi_{\rm nano}^{\rm lower} \lesssim 2$ instead of $\phi_{\rm
nano}^{\rm upper}/\phi_{\rm nano}^{\rm lower} \approx 10$.  Karanikas
and Louis suggest that these differences might stem from the
polydisperse nature of the nanoparticles employed in the experiments.
However, small-angle x-ray scattering measurements~\cite{martinez05}
have indicated that the nanoparticles, while certainly not monodisperse,
have a size distribution that is considerably narrower than indicated in
Ref.~\cite{tohver01} ($2{\rm nm} < \sigma_{\rm nano} < 4{\rm nm}$ rather
than $0.5{\rm nm} < \sigma_{\rm nano} < 11{\rm nm}$).  Nevertheless,
these differences are not surprising because Karanikas and Louis opted
not to include an explicit microsphere--nanoparticle attraction.
Accordingly, halo formation can only result from nanoparticle repulsion.
However, as pointed out in Ref.~\cite{liu04b} and in
Sec.~\ref{sec:intro} of the present paper, this mechanism is highly
unlikely to be relevant under the conditions of Ref.~\cite{tohver01},
and can indeed only be made to work at either larger screening lengths
or higher nanoparticle concentrations.

While, as shown above and in Ref.~\cite{liu04b}, the presence of an
induced, short-range colloid--nanoparticle attraction can resolve this
inconsistency, the inclusion of an attractive
interaction~$V_{\text{m-n}}$ yields rather different results for the
screening length chosen in the integral-equation
approach~\cite{karanikas04}. If $V_{\text{m-n}}$ and the nanoparticle
repulsion~$V_{\text{nano}}$ have a comparable range, bridging
attractions arise that have a detrimental effect on colloidal stability.
On the other hand, if $V_{\text{m-n}}$ is relatively long-ranged
compared to $V_{\text{nano}}$, bridging is avoided, but also the
reentrant gelation is no longer observed for the entire range of
nanoparticle volume fractions investigated~\cite{karanikas04}.

\begin{figure}
  \centering \includegraphics[width=\figurewidth,angle=0]{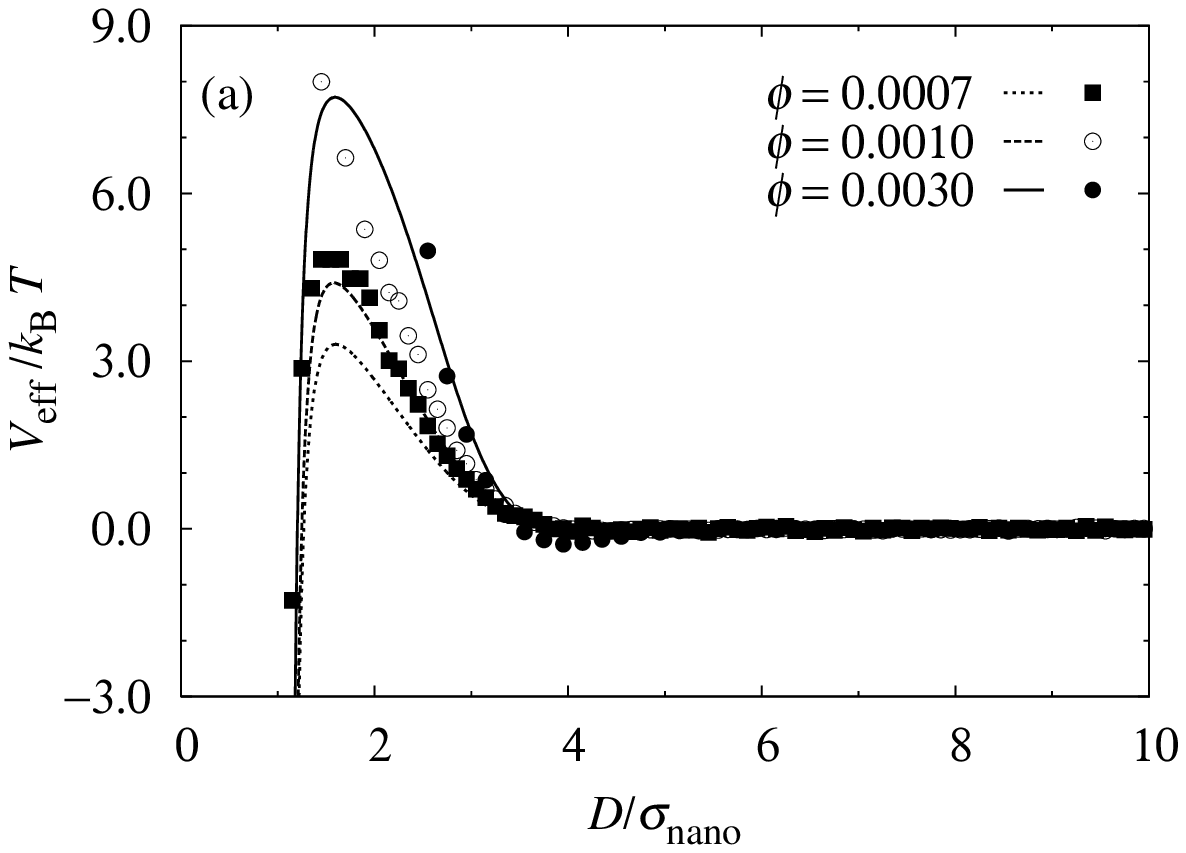}
  \includegraphics[width=\figurewidth,angle=0]{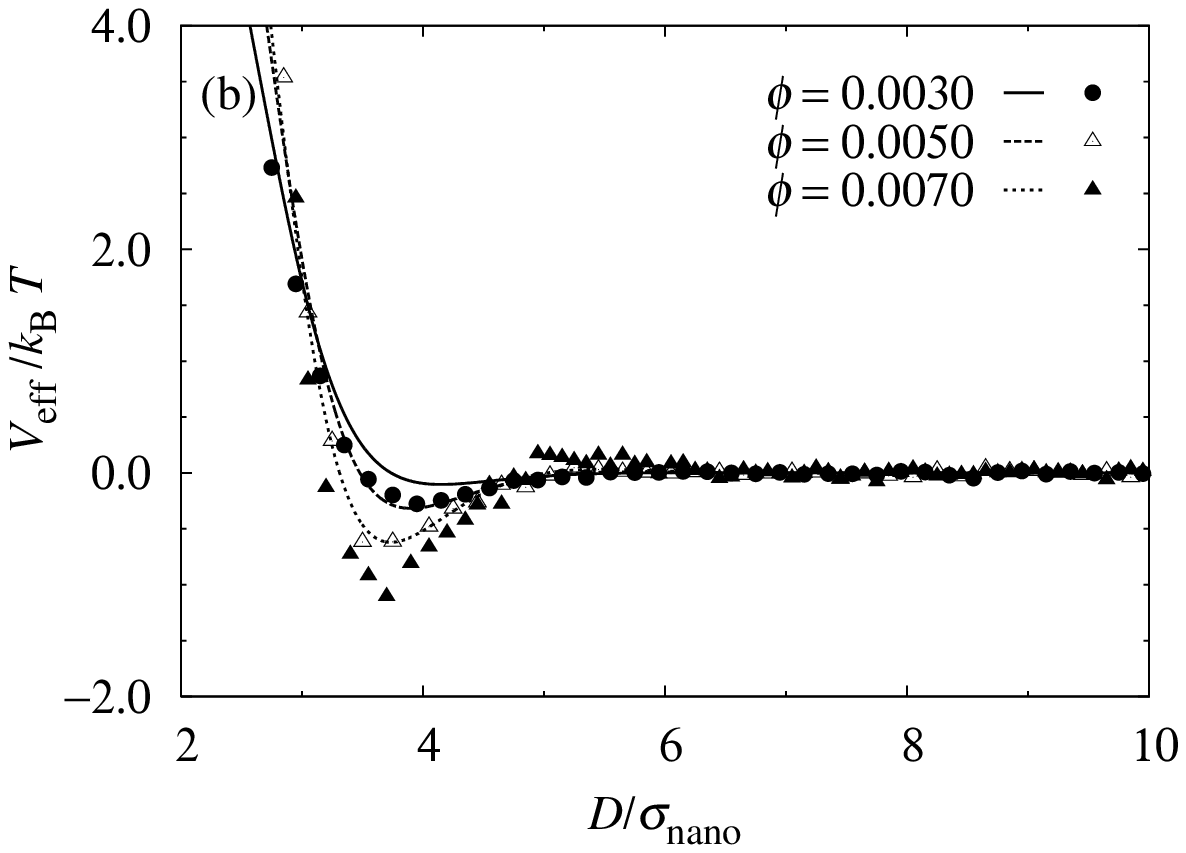}
  \caption{Comparison of the Monte Carlo results of Figs.\
  \ref{fig:low}(b) and~\ref{fig:high}(b) (symbols) to HNC
  integral-equation results (lines) \cite{karanikas05}.  (a) At low
  nanoparticle volume fractions, the analytical results reproduce the
  nanoparticle bridging at $D=\sigma_{\rm nano}$ as well as the
  repulsive barrier. Although the height of this barrier is
  underestimated, it increases with increasing nanoparticle
  concentration, as in the simulations.  (b) At higher nanoparticle
  volume fractions, the integral-equation results reproduce the
  attractive minimum found in Fig.~\ref{fig:high}(b). The strength of
  the attraction is approximately twice smaller than in the MC
  calculations.}  
  \label{fig:louis}
\end{figure}

It is also of interest to consider instead the predictions of an
integral-equation approach for the pair potentials employed in the Monte
Carlo simulations~\cite{liu04b}. Figure~\ref{fig:louis} presents a
comparison between the simulation results of Figs.\ \ref{fig:low}(b)
and~\ref{fig:high}(b) and corresponding integral-equation results
obtained using the HNC closure~\cite{karanikas05}. Despite the fact that
HNC performs best for smoothly varying potentials~\cite{hansen86}
(whereas the interactions in Sec.~\ref{sec:interaction} vary rapidly at
short separation), the integral-equation results display
semi-quantitative agreement with our simulation results. For low
$\phi_{\rm nano}$ [Fig.~\ref{fig:louis}(a)], both the bridging
attraction at $\sigma_{\rm nano}$ and the strong effective repulsion for
larger microsphere separation are reproduced, although the height of the
barrier is systematically underestimated. At $\phi_{\rm nano}=0.0007$
the maximum in $V_{\rm eff}$ is approximately 40\% lower than in the MC
calculations.  For higher nanoparticle concentrations
[Fig.~\ref{fig:louis}(b)], the secondary minimum near $D =
4.0\sigma_{\rm nano}$ is also recovered in the HNC calculations, albeit
at a weaker strength than the simulations. Overall, we view this as a
significant confirmation of the results presented in Ref.~\cite{liu04b}.

\section{Summary and conclusions}

We have presented a detailed numerical study of effective interactions
between micron-sized silica spheres, induced by highly charged zirconia
nanoparticles. Our calculations provide an explanation for the
nanoparticle haloing phenomenon discovered by Lewis and
co-workers~\cite{tohver01}, reproducing both colloidal stabilization and
reentrant gelation in quantitative agreement with the experimental
findings.  The haloing mechanism for colloidal stabilization is found to
rely on rather generic features and hence may become of considerable
practical importance. Our calculations provide explicit guidance for the
role of charge and size asymmetry between the colloids and the
nanoparticles. The presence of weak colloid--nanoparticle attractions is
found to be a crucial ingredient for stabilization at the very low
nanoparticle volume fractions and strong screening conditions in the
experiments.  Integral-equation calculations using the hypernetted-chain
closure have provided a semi-quantitative confirmation of the effective
interactions computed in the Monte Carlo simulations. We note the
simulations involve not only large size asymmetries, but also extremely
large numbers of particles. The use of a novel geometric Monte Carlo
algorithm~\cite{dress95,geomc,liu05a} was therefore indispensable.

\begin{acknowledgments}
  We gratefully acknowledge stimulating discussions with Jennifer Lewis
  and Ken Schweizer. We thank Ard Louis and Stelios Karanikas for
  generously permitting us to include the integral-equation results of
  Ref.~\cite{karanikas05} prior to publication.  This material is based
  upon work supported by the National Science Foundation under CAREER
  Award No.\ DMR-0346914 and Grant No.\ CTS-0120978 and by the U.S.
  Department of Energy, Division of Materials Sciences under Award No.\
  DEFG02-91ER45439, through the Frederick Seitz Materials Research
  Laboratory at the University of Illinois at Urbana-Champaign.  The
  calculations presented here were in part performed on the NSF TeraGrid
  facility at the National Center for Supercomputing Applications under
  Award No.  TGDMR040030T.
\end{acknowledgments}

\end{document}